\documentclass[notoc,12pt]{JHEP3}

\usepackage{mcite}
\usepackage{graphicx}
\usepackage{amsmath}

\newcommand{\graphicswidth}{0.6\textwidth}
\newcommand{\dd}{\mathrm{d}}

\title{
From Tree Unitarity to Top Quark Physics
in 5D Higgsless Models\thanks{Talk given at the
International Europhysics Conference On High Energy Physics, July 21st -27th 
2005, Lisboa, Portugal}
}

\author{Christian Schwinn\thanks{Work supported by by the Deutsche Forschungsgemeinschaft through the
Gra\-du\-ier\-ten\-kolleg `Eichtheorien' at Mainz University}\\
         Institut f\"ur Physik, Johannes-Gutenberg-Universit\"at\\
 Staudingerweg 7,  D-55099 Mainz, Germany\\
        E-mail: \email{schwinn@thep.physik.uni-mainz.de}}

      \abstract { 
        
        In five dimensional models of Higgsless electroweak symmetry
        breaking, tree level unitarity in gauge boson scattering is
        restored by the exchange of gauge boson Kaluza-Klein modes
        instead of a Higgs boson.  Unitarity of scattering amplitudes
        involving top quarks requires also the
        Kaluza-Klein modes of the third family quarks.
        It is shown that the 
        relevant unitarity cancellations are consistent with gauge
        symmetry breaking by boundary conditions.
        These results are used to constrain the couplings of the top quark to
        Kaluza-Klein modes and the implications for collider
        phenomenology are discussed.

}

\begin{document}

\section{Introduction: Tree unitarity in Higgsless models}
One of the motivations for expecting new physics at the TeV
scale is 
the violation of perturbative unitarity---implying the onset of
strong interactions---in the scattering of massive gauge bosons at the scale
$\Lambda\sim 1$-$2$ TeV in a theory without a Higgs boson or other new
particles.  Tree level unitarity in $W^+ W^-\to W^+ W^-$ scattering
can be restored
by introducing scalars $H$ or new vector bosons $V$ whose couplings satisfy the
 \emph{unitarity sum rule} (SR)
      \begin{equation}
\label{eq:sr}
 4m_W^2 g_{WW\gamma}^2 + (4m_W^2-3m_Z^2)g_{WWZ}^2 = 
       \sum_{H} g_{WWH}^2 +
            \sum_{V}(3m_{V}^2-4m_W^2)g_{WWV}^2     
     \end{equation}
The conventional solution to such SRs---realized in the
Higgs mechanism---introduces one ore more scalar particles and no
additional  vector bosons.
The alternative possibility of
additional vector bosons and no scalar
is realized in a 
Kaluza-Klein~(KK)  gauge theory
and has led to the construction of higher dimensional 
electroweak Higgsless models~\cite{Csaki:2003dt}. With the lightest
KK-modes at $500$-$700$ GeV, these models remain
weakly coupled up to a  cutoff of $5$-$10$ TeV implied by partial wave 
unitarity~\cite{SekharChivukula:2001hz,*Papucci:2004ip}.
The gauge symmetry is broken 
 at the boundaries of the  extra dimension
 by assigning Dirichlet 
boundary conditions~(BCs) to the broken gauge fields
while the unbroken ones satisfy Neumann BCs.
The consistency of these BCs
with unitarity~\cite{Csaki:2003dt,OS:SR} 
is discussed in section~\ref{sec:bcs}.
The collider signatures of such models have been 
constrained using the unitarity SRs~\cite{Birkedal:2004au}, 
resulting in  a bound 
$g_{ZWW^{(1)}}\lesssim \frac{g_{ZWW}m_Z^2}{\sqrt 3 m_{W^{(1)}} m_W}$  
on the coupling of the first KK-mode of the $W$. 
In section~\ref{sec:fermi} the implications of a similar analysis in the
top sector~\cite{CS:HLTOP} for
collider phenomenology are discussed.

\section{Consistency of boundary conditions in 5D gauge theories}
\label{sec:bcs}
Consistency of the
Becchi-Rouet-Stora-Tuytin (BRST) symmetry can be used as
a simple criterion
for  the compatibility of symmetry breaking by BCs
with the unitarity SRs~\cite{OS:SR}.
The KK wavefunctions of the 4D components of the gauge bosons
$A^{a,(n)}_{\mu}(x)$ will be denoted by $f^a_{n}(y)$ and those of the
fifth component $A^{a,(n)}_{5}(x)$ by $g^a_{n}(y)$ 
where one can impose the relation
$\partial_y f^a_n
=m_{A^{a,(n)}}g^a_n$. 
The KK-modes of $A_5$
play the role of Goldstone bosons with the BRST-transformation 
$ \delta_{\text{BRST}} A_5^\alpha=
m_{A^\alpha} c^\alpha
     +T^{\gamma}_{\alpha\beta} A_5^{\beta}c^{\gamma}$
where a multi-index $\alpha=(a,n)$ is used.
The constants $T^{\gamma}_{\alpha\beta}$ also enter the Lagrangian:
\begin{equation}
   \mathcal{L}_{KK} =
    - g^{\alpha\beta\gamma}
    \partial_\mu A_\nu^\alpha
 A^{\beta,\mu}A^{\gamma,\nu} 
    - \frac{1}{2}T^{\alpha}_{\beta\gamma}\, A^{\alpha,\mu}
      A_5^{\beta}\overleftrightarrow{\partial_\mu} 
      A_5^\gamma +\dots
 \end{equation}
The coupling constants are given in terms of the structure constants
$f^{abc}$ and the KK-wavefunctions:
 \begin{equation}
\label{eq:couplings}
  g^{\alpha \beta\gamma}
    = f^{abc}\int\!\dd y\, f^\alpha (y)
    f^\beta(y) f^\gamma (y) \quad,\quad
  T^{\alpha}_{\beta\gamma}
    = f^{abc}\int\!\dd y \, f^\alpha (y) g^\beta(y)
    g^\gamma (y)
\end{equation}         
Requiring nilpotency of the BRST transformation, i.e.
$ \delta_{\text{BRST}}^2 A_5^\alpha=0$, results in the
conditions
 \begin{equation}
\label{eq:nil}
 m_{A^\beta}T^\gamma_{\alpha \beta}
  -m_{A^{\gamma}}T^{\beta}_{\alpha\gamma}
  =m_{A^\alpha} g^{\alpha\beta\gamma} \quad,\quad
T^{\gamma}_{\alpha\beta}
 T^{\delta}_{\beta\epsilon}-
 T^{\delta}_{\alpha\beta}
   T^{\gamma}_{\beta\epsilon}=
  g^{\beta\delta\gamma}T^{\beta}_{\alpha\epsilon}
 \end{equation}  
where the transformation of the ghost fields 
$\delta_{\text{BRST}} c^\alpha = \frac{1}{2}g^{\alpha\beta\gamma}c^\beta c^\gamma$ 
has been used in addition. 
Inserting the explicit expressions~\eqref{eq:couplings},  
the first condition in~\eqref{eq:nil} involves the values of the
KK-wavefunctions on the boundaries, and thus gives a criterion
for the consistency of BCs:
\begin{equation}
\label{eq:bc-consistency}
0=f^{abc}\int\!\dd y\,\partial_y(g^\alpha f^\beta f^\gamma)
      =f^{abc}\left(g^\alpha f^\beta f^\gamma\right)
      \Bigr|^{\pi R}_0
\end{equation}
One can show~\cite{OS:SR}
 that~\eqref{eq:bc-consistency} is satisfied for Neumann BCs
$\partial_y f^\alpha|_{0,\pi R} =0= g^\alpha|_{0,\pi R}$  
and for  
Dirichlet BCs $ f^\alpha|_{0,\pi R} =0=\partial_y
 g^\alpha|_{0,\pi R}$ as
used for symmetry breaking in 5D Higgsless
models. Using the solution of the first condition, 
the second condition in~\eqref{eq:nil} gives
the unitarity SR for the cancellation of terms growing like
$E^2$ and can be verified using the completeness relations of the
KK-wavefunctions. Therefore the unitarity SRs are not spoiled by symmetry 
breaking by BCs.
\section{Unitarity sum rules for fermions and implications for top quark physics}
\label{sec:fermi}
Fermion masses in 5D Higgsless models
can be generated by gauge invariant brane localized mass and kinetic terms
for bulk fermions~\cite{Csaki:2003sh} that respect
the relevant unitarity SRs~\cite{CS:HLF}.
Electroweak precision data require  to delocalize the zero modes of light
 fermions so they decouple from
KK gauge bosons~\cite{Cacciapaglia:2004rb} and 
collider signatures are expected to be limited to the third family.
Realistic values of the $t$ and $b$ quark masses
 require their first KK-modes 
to be heavier than those of the $W$ and $Z$~\cite{Cacciapaglia:2004rb}. 
This 
has been achieved recently in
theory space inspired models~\cite{Georgi:2005dm,Foadi:2005hz}. 
Following~\cite{Birkedal:2004au},  
unitarity SRs can be used to constrain the interactions
of the KK-modes with the top-and bottom quarks~\cite{CS:HLTOP}.
The SRs for the process $W^+ W^-\to t\bar t$ 
result in the condition~\cite{CS:HLTOP}
\begin{equation}
             (g^{L}_{Wtb})^2= \sum_{n}\left[ 2\tfrac{m_{B^{(n)}}}
               {m_t}g^R_{WtB^{(n)}}g^L_{WtB^{(n)}}
     -(g^{L}_{WtB^{(n)}})^2-(g^{R}_{WtB^{(n)}})^2\right]
           \end{equation}
that has been recently verified in a concrete model~\cite{Foadi:2005hz}.
This condition shows that the KK-modes of the bottom quark are necessary for
the unitarity cancellations.
 Taking also
processes like $Z Z\to t\bar t$ and $Z W^+\to t\bar b$ into account
and truncating after the first KK-level, 
 the unitarity SRs lead to
the estimates~\cite{CS:HLTOP}
$g_{Wt B^{(1)}}
\approx \tfrac{g}{2}\sqrt{\tfrac{ m_t}{m_{B^{(1)}}}}$ and 
$g_{tt Z^{(1)}}\approx
\frac{\sqrt 3 g}{4}\frac{m_t m_{Z^{(1)}}}{m_Wm_{B^{(1)}} }$.
As seen in figure~\ref{fig:wwtt}, this choice of couplings 
suppresses the $W^+ W^-\to t\bar t$ cross section at high energies.
\FIGURE{
   \includegraphics[width=\graphicswidth]{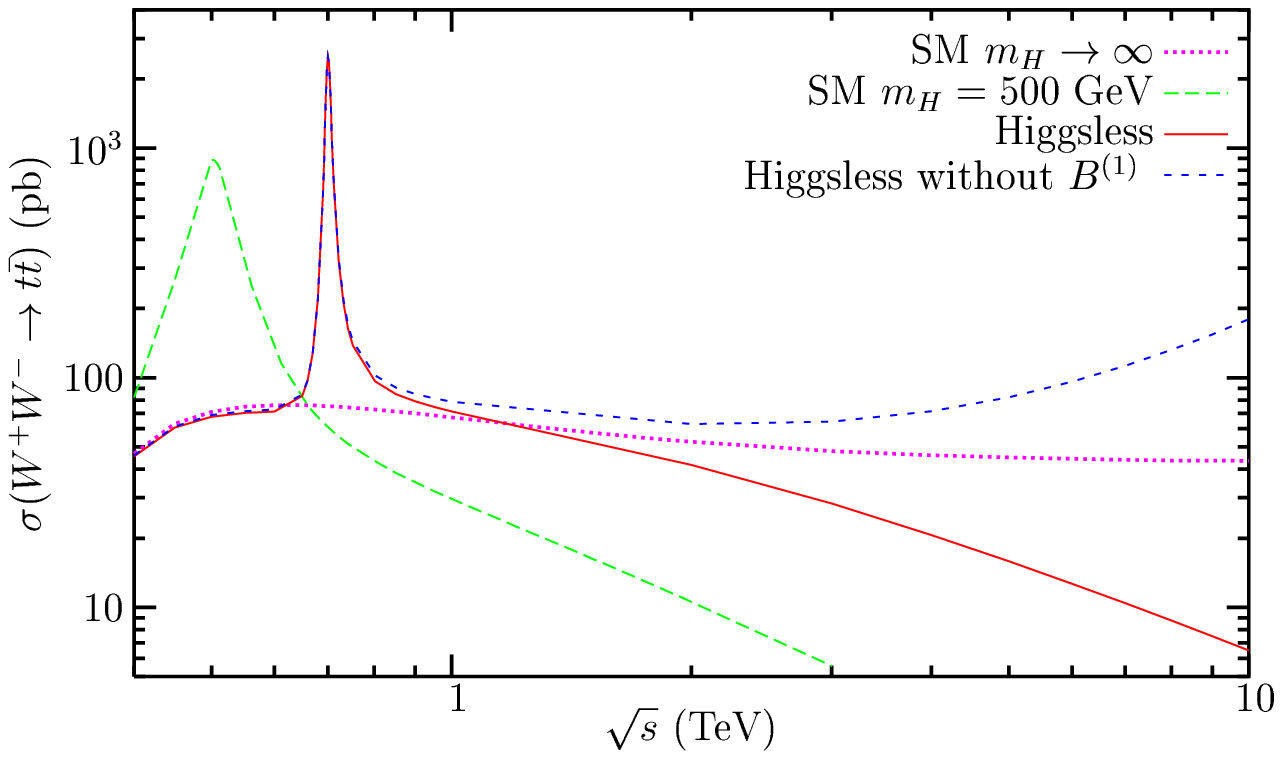}
    \caption{Cross section for $W^+ W^-\to t\bar t$ in the
      SM with a Higgs resonance, the SM in the $m_H\to\infty$
      limit, the Higgsless scenario (with $m_Z^{(1)}=700$ GeV and
      $m_{B^{(1)}}=2.5$ TeV) and the Higgsless scenario without $B^{(1)}$}
\label{fig:wwtt}
}

The above results can be used~\cite{CS:HLTOP} to estimate 
the $ Z^{(1)}\to t\bar t$ partial width to be
about $14\%$ of the $Z^{(1)}\to W^+W^-$ partial
 width for $m_{B^{(1)}}=2.5$ TeV,
while it grows to $86\%$ for $m_{B^{(1)}}=1$ TeV.
The most useful LHC signatures of a neutral gauge
boson coupling mainly to third generation quarks 
are the associated
production~\cite{Han:2004zh} in processes like
 $W b\to t Z^{(1)}$ or  $g g \to t\bar t Z^{(1)}$. 
At an $e^+ e^-$ linear collider
such a $Z$ resonance 
can be probed in the
 vector boson fusion process $W^+ W^- \to \bar t t$~\cite{Han:2000ic}. 

The properties of the $T^{(1)}$ can be compared to those of the heavy top
in the littlest Higgs model~\cite{Han:2003wu}. 
The decay widths in the two scenarios are related by
$ \Gamma^{\text{HL}}_{T^{(1)}\to t Z}
\approx \frac{m_T}{4 m_t}  \Gamma_{T}^{\text{LH}}$
so in the Higgsless scenario the $T^{(1)}$ is broader,
 for instance $m_{T^{(1)}}=3$ TeV 
implies $\Gamma_{T^{(1)}\to t Z}\approx 270$
GeV compared to $\Gamma_T^{LH}=62$ TeV in the
little Higgs model. 
The discovery reach of the LHC in $W$-$b$ fusion $q b\to q' T$
has been estimated~\cite{Han:2003wu} as $m_T=2$ TeV.
The production cross section in the two scenarios
can be related  by 
$ \sigma_{\text{HL}}(Wb\to T^{(1)})\approx
 \frac{ m_b m_T^{(1)}}{m_t^2} 
 \sigma_{\text{LH}}(Wb\to T) $
so the detection of the $T^{(1)}$ in the Higgsless
model is expected to be more challenging than in the Littlest Higgs model.

\section{Conclusions}
 I have described theoretical methods to verify unitarity
sum rules in higher dimensional Higgsless models and the phenomenological
consequences in the top quark sector.
As discussed in section~\ref{sec:bcs}, gauge symmetry breaking by boundary
conditions used in these models is compatible with  BRST
symmetry, as required  for a consistent quantization.
Unitarity sum rules have been used in section~\ref{sec:fermi}
to constrain the couplings of the KK-modes of the third family quarks and
gauge bosons to the top quark and the
implications for collider phenomenology have been discussed.

\bibliography{biblio}

\end{document}